\documentclass[12pt]{article}
\usepackage{epsfig,amssymb}
\usepackage{latexsym}

\newcommand{\bq}{\begin{equation}}
\newcommand{\eq}{\end{equation}}
\newcommand{\bqa}{\begin{eqnarray}}
\newcommand{\eqa}{\end{eqnarray}}
\newcommand{\ben}{\begin{enumerate}}
\newcommand{\een}{\end{enumerate}}
\newcommand{\bc}{\begin{center}}
\newcommand{\ec}{\end{center}}
\newcommand{\bqb}{\begin{eqnarray*}}
\newcommand{\eqb}{\end{eqnarray*}}

\def\gsim{\gtrsim}

\def\np#1#2#3{ Nucl. Phys. ${\bf{#1}}$:#2 (#3)}

\def\ie{{\it i.e. ~}}
\def\eg{{\it e.g. ~}}

\global\nulldelimiterspace = 0pt

\begin{document}
\pagenumbering{arabic}
\thispagestyle{empty}
\def\thefootnote{\fnsymbol{footnote}}
\setcounter{footnote}{1}

\begin{flushright}
November 7, 2007 \\
PTA/07-052\\
ArXiv:0711.     /[hep-ph]\\

 \end{flushright}
\vspace{2cm}
%---------------------titre ---------------------------------------
\begin{center}
{\Large\bf Subleading terms in asymptotic\\
 \vspace{0.2cm}
Passarino-Veltman functions}
 \vspace{1.5cm}  \\
%-----------------------------------------------------------------
{\large M. Beccaria$^{a,b}$, G.J. Gounaris$^c$, J. Layssac$^d$,
and F.M. Renard$^d$}\\
\vspace{0.2cm}
$^a$Dipartimento di Fisica, Universit\`a del Salento \\
Via Arnesano, 73100 Lecce, Italy.\\
\vspace{0.2cm}
$^b$INFN, Sezione di Lecce\\
\vspace{0.2cm}
$^c$Department of Theoretical Physics, Aristotle
University of Thessaloniki,\\
Gr-54124, Thessaloniki, Greece.\\
\vspace{0.2cm}
$^d$Laboratoire de Physique Th\'{e}orique et Astroparticules,
UMR 5207\\
Universit\'{e} Montpellier II,
 F-34095 Montpellier Cedex 5.\\
\end{center}

\vspace*{1.cm}
\begin{center}
{\bf Abstract}
\end{center}

We write explicit and self-contained asymptotic expressions
for the tensorial $B$, $C$ and $D$ Passarino-Veltman functions. These
include quadratic and  linear logarithmic terms,  as well as subleading
constant terms. Only mass-suppressed ${\cal O}(m^2/s)$ contributions
are neglected. We discuss the usefulness of such expressions,
 particularly for studying one-loop
effects in  2-to-2 body processes at high energy.

\vspace{0.5cm}

\def\thefootnote{\arabic{footnote}}
\setcounter{footnote}{0}
\clearpage

\section{Introduction}

  The relevance of high energies for simplifying the
   parameter independent  tests of the
Standard Model (SM) and its supersymmetric extensions,
like \eg  the Minimal Supersymmetric
Standard Model (MSSM), has been stressed during the last
few years with applications to lepton and hadron colliders.
Indeed, at the one-loop leading logarithmic level, the high energy behavior
of the various 2-body helicity amplitudes in $e^+e^-$,
$\gamma\gamma$, $q\bar q$, $qg$, $gg$,... processes reflects
in a direct  way the gauge and the Yukawa structures
of the basic Lagrangian.
For reviews in SM see \cite{SMrules}; while
the MSSM case has been studied  in  \eg \cite{MSSMrules}, where
  the leading  1-loop  SUSY and standard virtual  effects have been identified
  for  processes containing   any kind of  external  particles.
In particular, simple rules have been established giving the coefficients
of the leading $\ln$ and $\ln^2$ contributions.

 These rules
can be checked explicitly by computing
 the one loop diagrams in terms of Passarino-Veltman (PV)
functions \cite{Veltman}, and then using their asymptotic expansion
at the leading logarithmic level (LL).  Such calculations have
 already appear  in \cite{SMrules, MSSMrules}.\\

However, things are less simple if one wants to keep
the  subleading no-logarithmic asymptotic contributions, described by the so called
{\em constant} terms, which  are independent of the invariant c.m. energy
of any pair of   external legs.
Such  terms include a priori   true constants (numbers), but also
angular dependent contributions, or terms involving
ratios of external and/or internal masses.
The later are particularly relevant for  SUSY cases,
involving  diagrams  containing many different internal masses.
In such cases, the {\em constant}  subleading terms may be used to identify
tests of   model-parameters at high energies,
 which may be  simpler than whatever is possible at lower energies.

Another possibility is to explicitly check  the details of the
remarkable total helicity conservation (HC) property, which has been
established to all orders,  for any 2-to-2 body process
at high energy $(s,~|t|,~|u|)$,  in  any  supersymmetric
extension of the standard model \cite{heli}.
According to HC, in such supersymmetric extensions, only the amplitudes where
{\it the sum of the helicities of
the two incoming particles equals the sum of the helicities
of the two outgoing ones}, could be non-vanishing at
high energies and fixed angles.     \\

For these purposes it should be  convenient
to have at our disposal asymptotic  expressions of
the PV functions, which include also
the subleading terms contributing to the  {\em constants}  of
the physical asymptotic amplitudes discussed above. These  go  beyond
the  expressions   presented in \cite{Denner-Roth}, which only
include the logarithmic structure.
In achieving this we only need  the expressions for the $B_j$ functions
and the tensorial decomposition of  $C_j$ and $D_j$ functions partly given in
  \cite{Hag}, combined with  the asymptotic expressions of
the $C_0$ and $D_0$ functions
established by Denner and Roth in \cite{Denner-Roth}.

In the present work we use the same notation as in Hagiwara et al \cite{Hag}.
For convenience, we thought that it would be worthwhile
to include in the paper all aforementioned formulae.
We hope this  will be useful for future analyses
of the kind  mentioned above.

The contents of the paper are as follows.
In Sect. 2 we write   the definitions
for the $C_j$ and $D_j$  functions, as well as the exact expressions for the
and $B_j$ functions. In Sect.3 we give the explicit
analytical results for the asymptotic quadratic and linear logarithmic
terms involving the correct mass-scales, as well as the
 subleading {\it constant} terms of $B$, $C$ and $D$ functions.
In Sect. 4 we present some illustrations and discuss
specific properties. The conclusions are given in Section 5; while
 in the appendices we present    the connections between the  the Hagiwara
notation and   the one adapted by  LoopTools  \cite{Hag, looptools},
as well the  reduction formalism for the $C_j$ and $D_j$ functions. \\

\section{Definitions and conventions}

This writing is self-contained, meaning that all definitions
and conventions have been recalled in a uniform
fashion in terms of external and internal masses
and momenta. We use the  Hagiwara  et al \cite{Hag} definitions
of the tensorial functions, but in Appendix
we give also the relations with LoopTools definitions
\bqa
&& A(m_1)={(2\pi\mu)^{2\epsilon}\over i\pi^2}
\int {d^nk\over N_1}
=m^2_1\left (\Delta -\ln\frac{m^2_1}{\mu^2}+1 \right ) ~~,  \label{A-exact} \\
&& [B_0,B^{\mu},B^{\mu\nu}](12)
={(2\pi\mu)^{2\epsilon}\over i\pi^2}
\int {d^nk~[1,k^{\mu},k^\mu k^\nu]\over N_1N_2} ~~, \label{B-int} \\
&& [C_0,C^{\mu},C^{\mu\nu},C^{\mu\nu\rho}](123)=
{(2\pi\mu)^{2\epsilon}\over i\pi^2}
\int {d^nk~[1,k^{\mu},k^\mu k^\nu,k^\mu k^\nu k^\rho]
\over N_1N_2N_3} ~~,  \label{C-int}\\
&& [D_0,D^{\mu},D^{\mu\nu},D^{\mu\nu\rho}](1234)=
{(2\pi\mu)^{2\epsilon}\over i\pi^2}
\int {d^nk~[1,k^{\mu},k^\mu k^\nu,k^\mu k^\nu k^\rho]
\over N_1N_2N_3N_4} ~~, \label{D-int}
\eqa
with
\bq
n=4-2\epsilon~~~,~~~\Delta={1\over\epsilon}-\gamma_E+\ln(4\pi) ~~, \label{n-Delta}
\eq
\bqa
&& N_1=k^2-m^2_1+i\epsilon ~~, \nonumber \\
&& N_2=(k+p_1)^2-m^2_2+i\epsilon ~~, \nonumber  \\
&& N_3=(k+p_1+p_2)^2-m^2_3+i\epsilon ~~, \nonumber \\
&& N_4=(k+p_1+p_2+p_3)^2-m^2_4+i\epsilon ~~. \label{N-den}
\eqa

\begin{figure}[ht]
%\vspace*{-3cm}
\[
\hspace{-0.5cm}\epsfig{file=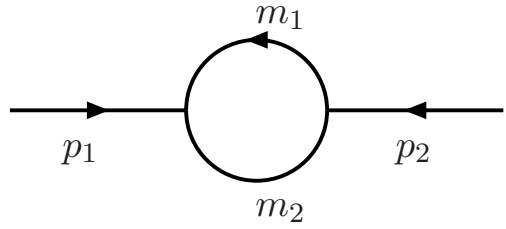,height=3.cm}
\]
\vspace*{-0.5cm}
\caption[1]{Bubble graph for $B_j(12)$. Of course $p_1^2=p_2^2$. }
\label{bub-fig}
\end{figure}

\begin{figure}[ht]
%\vspace*{-3cm}
\[
\hspace{-0.5cm}\epsfig{file=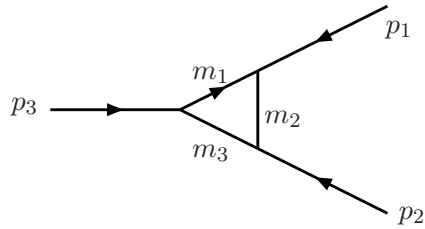,height=3.cm}
\]
\vspace*{-0.5cm}
\caption[1]{Triangular graph for  $C_j(123)$.  }
\label{tri-fig}
\end{figure}

\begin{figure}[b]
%\vspace*{-3cm}
\[
\hspace{-0.5cm}\epsfig{file=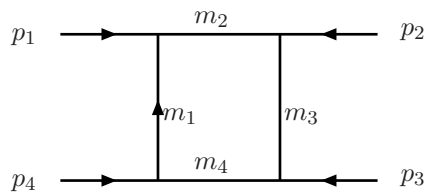,height=2.5cm}
\]
\vspace*{-1.cm}
\caption[1]{Box graph for $D_j(1234)$. }
\label{box-fig}
\end{figure}

The definitions of the above functions
are completed by  the Figures \ref{bub-fig}, \ref{tri-fig}, \ref{box-fig},
where all external momenta  are incoming,
   and the arrowed internal line carries  the momentum $k$
   in  the direction of the arrow; compare  (\ref{A-exact}-\ref{D-int}).
The internal masses are also indicated there.

Following Hagiwara  et al \cite{Hag},  we expand  the tensorials in
(\ref{B-int}, \ref{C-int}, \ref{D-int}) for the respective $B_j$, $C_j$ and $D_j$
functions using the definitions
\bqa
&& B^{\mu}(12)=p^{\mu}_1B_1(12) ~~, \nonumber \\
&& B^{\mu\nu}(12)=p^{\mu}_1p^{\nu}_1B_{21}(12)+g^{\mu\nu}B_{22}(12) ~~,
 \nonumber \\
&& B_j(12)=B_j(p_1^2;m_1,m_2)=B_j(p_2^2;m_1,m_2)~~,   \label{B-tensor}
\eqa
\medskip
\bqa
&& C^{\mu}(123)=p^{\mu}_1C_{11}(123)+p^{\mu}_2C_{12}(123) ~~, \nonumber  \\
&& C^{\mu\nu}(123)=p^{\mu}_1p^{\nu}_{1}C_{21}(123)
+p^{\mu}_2p^{\nu}_{2}C_{22}(123)
+(p^{\mu}_1p^{\nu}_{2}+p^{\mu}_2p^{\nu}_{1})C_{23}(123)
+g^{\mu\nu}C_{24}(123) ~~, \nonumber \\
&& C^{\mu\nu\rho}(123)=\sum_{i=1,2} C_{00i}(123)
(g^{\mu\nu}p^{\rho}_i
+g^{\nu\rho}p^{\mu}_i+g^{\mu\rho}p^{\nu}_i)
+\sum_{i,j,k=1,2} C_{ijk}(123)p^{\mu}_ip^{\nu}_jp^{\rho}_k ~~, \nonumber \\
&& C_j(123)=C_j(p_1^2, p_2^2, p_3^2; m_1, m_2, m_3)~~, \label{C-tensor}
\eqa
\medskip
\bqa
 D^{\mu}(1234)&=& p^{\mu}_1D_{11}(1234)+p^{\mu}_2D_{12}(1234)
+p^{\mu}_3D_{13}(1234 ~~, \nonumber \\
 D^{\mu\nu}(1234)&=&p^{\mu}_1p^{\nu}_{1}D_{21}(1234)
+p^{\mu}_2p^{\nu}_{2}D_{22}(1234)
+p^{\mu}_3p^{\nu}_{3}D_{23}(1234)\nonumber\\
&& +(p^{\mu}_1p^{\nu}_{2}+p^{\mu}_2p^{\nu}_{1})D_{24}(1234)
+(p^{\mu}_1p^{\nu}_{3}+p^{\mu}_3p^{\nu}_{1})D_{25}(1234) \nonumber \\
&& +(p^{\mu}_2p^{\nu}_{3}+p^{\mu}_3p^{\nu}_{2})D_{26}(1234)
+g^{\mu\nu}D_{27}(1234) ~~, \nonumber \\
D^{\mu\nu\rho}(1234)& =& \sum_{i=1,2,3} D_{00i}(1234)
(g^{\mu\nu}p^{\rho}_i +g^{\nu\rho}p^{\mu}_i+g^{\mu\rho}p^{\nu}_i)
+ \sum_{i,j,k=1,2,3} D_{ijk}(1234)p^{\mu}_ip^{\nu}_jp^{\rho}_k ~~, \nonumber \\
D_j(1234)&=& D_j(p^2_1,p^2_2,p^2_3,p^2_4,(p_1+p_2)^2, (p_2+p_3)^2;m_1,m_2,m_3,m_4)
~~. \label{D-tensor}
\eqa
Particularly for the $D$-functions in (\ref{D-tensor}), the notation
\bq
t=(p_1+p_2)^2~~,~~s=(p_2+p_3)^2~~, u=(p_1+p_3)^2~~, \label{Dkin1}
\eq
is also convenient; compare Fig.\ref{box-fig}. Since  $C_{ijk}$ and $D_{ijk}$
in (\ref{C-tensor}, \ref{D-tensor})
do not depend on the permutation of their indices,  they are
traditionally defined with the indices in ascending order.\\

In the case of $B$ functions,  the exact
expressions for any $s=p^2_1=p^2_2$,
may be obtained  by integrating (\ref{B-int}), which gives
\bqa
&& B_0(q^2;m_1,m_2)=\Delta -\ln\frac{m_1m_2}{\mu^2}+2 +
\frac{1}{q^2} \Big [ (m_2^2 -m_1^2)\ln\frac{m_1}{m_2}
\nonumber \\
&& + \sqrt{\lambda(q^2+i\epsilon, m_1^2, m_2^2)}
\, {\rm ArcCosh} \Big (\frac{m_1^2+m_2^2-q^2-i\epsilon}{2 m_1 m_2} \Big ) \Big ] ~~,
\nonumber \\
&& B_1(q^2;m_1,m_2)=\frac{1}{2q^2}\Big [ A(m_1)-A(m_2)
+(m_2^2-m_1^2-q^2)B_0(q^2;m_1,m_2)\Big ]
~~,  \nonumber \\
&& B_{21}(q^2;m_1,m_2)=\frac{1}{3q^2}\Big [A(m_2)- m_1^2B_0(q^2;m_1,m_2)
-2(q^2+m_1^2-m_2^2)B_1(q^2;m_1,m_2)
\nonumber \\
&& \hspace{3cm}  -\frac{(m_1^2+m_2^2)}{2}+\frac{q^2}{6} \Big ]
\nonumber \\
&& ~~
 B_{22}(q^2;m_1,m_2)=\frac{1}{6}\Big [A(m_2)+2 m_1^2
B_0(q^2;m_1,m_2)
+(m_1^2-m_2^2+q^2)B_1(q^2;m_1,m_2)
\nonumber \\
&& \hspace{3cm}  -\frac{q^2}{3}+m_1^2+m_2^2 \Big ] ~~,
\label{B-exact}
\eqa
where
\bq
\lambda(a,b,c)=a^2+b^2+c^2-2ab-2ac-2bc~~. \label{lambda-term}
\eq
and
\bq
{\rm ArcCosh} \Big (\frac{m_1^2+m_2^2-q^2-i
\epsilon}{2 m_1 m_2} \Big )=\ln (z+\sqrt{z^2-1}) ~~~,~ ~~
z=\left (\frac{m_1^2+m_2^2-q^2-i\epsilon}{2 m_1 m_2}\right )~~. \label{ArcCosh}
\eq

Separating out the divergent and $\mu$-dependent parts in (\ref{B-exact}),
we obtain
\bqa
B_0(q^2;m_1,m_2)&=&\Delta-\ln{m_1m_2\over\mu^2} + b_0(q^2;m_1,m_2)  ~~, \nonumber \\
B_1(q^2;m_1,m_2)&=& -\frac{1}{2}\left[\Delta-\ln{m_1m_2\over\mu^2}\right]
+ b_1(q^2;m_1,m_2)  ~~, \nonumber \\
B_{21}(q^2;m_1,m_2)&=&\frac{1}{3}\left[\Delta-\ln{m_1m_2\over\mu^2}\right]
+ b_{21}(q^2;m_1,m_2)  ~~, \label{BDelta-relation}
\eqa
with
\bqa
b_0(q^2; m_1,m_2)&=&
2 + \frac{1}{q^2} \Big [ (m_2^2 -m_1^2)\ln\frac{m_1}{m_2}\nonumber\\
&& + \sqrt{\lambda(q^2+i\epsilon, m_1^2, m_2^2)}  {\rm ArcCosh} \Big
(\frac{m_1^2+m_2^2-q^2-i\epsilon}{2 m_1 m_2} \Big ) \Big ] ~~, \nonumber \\
b_1(q^2 ; m_1,m_2)&=&  {1\over2q^2} \Big [ m^2_1-m^2_2
+(m^2_1+m^2_2)\ln\frac{m_2}{m_1}\nonumber +(m_2^2 -m_1^2-q^2)b_0(q^2 ;m_1,m_2)\Big ]
 ~~, \nonumber \\
b_{21}(q^2 ; m_1,m_2)&=&
\frac{(q^2+m^2_1-m^2_2)^2-q^2 m^2_1}{3 q^4}\,\, b_0(q^2 ;m_1,m_2)
 +{(m^2_2-m^2_1)\over6q^2}\left [3+{2(m^2_1-m^2_2)\over q^2}\right ]
\nonumber \\
&&  +{1\over3q^2}\left [ 2m^2_2+m^2_1 -{(m^4_2-m^4_1)\over q^2}\right ]
\ln {m_1\over m_2} +\frac{1}{18}
~~. \label{b-exact}
\eqa
Using these, we define
\bqa
 &&  b_j^{(12)}\equiv b_j(p^2_1;m_1,m_2) ~~, \nonumber \\
&&  b_j^{(13)}\equiv b_j((p_1+p_2)^2;m_1,m_3)~~, \nonumber \\
&&  b_j^{(14)}\equiv b_j((p_1+p_2+p_3)^2;m_1,m_4) ~~, \nonumber  \\
&&  b_j^{(23)}\equiv b_j(p^2_2;m_2,m_3) ~~, \nonumber  \\
&&  b_j^{(24)}\equiv b_j((p_2+p_3)^2;m_2,m_4)~~, \nonumber \\
&&  b_j^{(34)}\equiv b_j(p^2_3;m_3,m_4) ~~, \label{bij-mass-ratios}
\eqa
which are in the same sprit as the expressions  (D.33) in \cite{Hag}.
These functions only depend on ratios of internal or external masses
and contribute to the {\em constant} asymptotic
terms  discussed above.

In Appendix 1 we give the relation of the above  expansion  for
the $B_j$, $C_j$ and $D_j$ functions,
with  that of the LoopTools library. From this expansion
one obtains the exact
expressions of the various PV functions  in terms
of the basic $B_0$, $C_0$, $D_0$ and of $A(m_i)$
for various combinations of external and internal
masses. Most of these relations have been written
in Hagiwara's Appendix D \cite{Hag};
in our Appendix 2 we have added a few
more, relevant   for the $C_{ijk}$ and $D_{ijk}$ functions. \\

\section{Asymptotic expansion of the B, C and D functions}

We consider non zero external
and internal masses. For what concerns very light leptons
and quarks, although logarithmic singularities
$\ln(s/m^2)$ may appear temporarily inside certain
PV functions, they finally disappear in physical amplitudes.
This can be checked explicitly in each particular
process, as it is just the consequence
of the theorem in \cite{sing}.

For what concerns the photon, we  keep a fictitious mass
$m_{\gamma}$. It can be used as an infrared regulator
which will finally disappear when adding the soft
photon radiation. \\

\subsection{Asymptotic B  functions}

Using (\ref{BDelta-relation}, \ref{b-exact}) and Fig.\ref{bub-fig},
at asymptotic energies $p_1^2\equiv s  \gg m^2_i$, one  gets
\bqa
 B_{0}(s;m_1,m_2)
& \simeq &\Delta+2-\ln s_\mu   ~~\nonumber \\
 B_1(s;m_1,m_2)& \simeq & - \frac{\Delta}{2}-1+\frac{\ln s_\mu}{2} ~~ \nonumber \\
 B_{21}(s;m_1,m_2) & \simeq &\frac{-\ln s_\mu}{3}
+\frac{\Delta}{3}+\frac{13}{18}  ~~, \nonumber \\
 B_{22}(s;m_1,m_2) & \simeq & \frac{s \ln s_\mu }{12}
-\frac{s\Delta }{12}-\frac{ 2 s}{9} ~, \label{B-asym}
\eqa
where the  definition
\bq
s_{\mu}\equiv \frac{-s -i\epsilon}{\mu^2}~~, \label{smu}
\eq
correctly describes the real and imaginary parts
at  asymptotic (positive or negative) $s$.
 We note that the asymptotic expressions   (\ref{B-asym}) contain only the
  leading logarithmic  and  subleading constant contributions,
  while the neglected terms  are  ${\cal O}(m_i^2/s)$.
The divergent quantity $\Delta$ has been defined in (\ref{n-Delta}).

\subsection{Asymptotic C  functions}

Based on Fig.\ref{tri-fig}, we   consider the case in which  the square
of only one of the external momenta is
large, the other two, as well as the internal masses, being much smaller; i.e.
\bq
p_3^2\equiv s=(p_1+p_2)^2 \gg (m_i^2~,~ |p_1^2| ~,~ |p_2^2|)~~.
\eq
Defining also
\bq   s_\mu =\frac{-s-i\epsilon }{\mu^2}~~,~~
 s_2=\frac{-s -i\epsilon}{m_2^2}~~,~~
s_{ij}=\frac{-s -i\epsilon}{m_im_j}~,  ~
\label{analytic-s12}
\eq
and neglecting terms like $(1/s){\cal O}(m_i^2/s)$,
the asymptotic expression of the $C_0$
function  established by Denner and Roth \cite{Denner-Roth} is written as
\bq
\label{c0expansion}
 C_0(p_1^2,p_2^2, s; m_1, m_2, m_3)\simeq
 \frac{(\ln s_2 )^2}{2 s}+\frac{L_{223}+L_{121}}{s},
\eq
where
\bqa
 L(p_a, m_b, m_c)\equiv L_{abc} & = &
 \phantom{+} {\rm Li_2} \left ( \frac{2p_a^2+i\epsilon}{m_b^2-m_c^2+p_a^2+i\epsilon +
\sqrt{\lambda (p_a^2+i\epsilon, m_b^2, m_c^2)}} \right )
\nonumber \\
&& + {\rm Li_2} \left ( \frac{2p_a^2+i\epsilon }{m_b^2-m_c^2+p_a^2+i\epsilon -
\sqrt{\lambda (p_a^2+i\epsilon, m_b^2, m_c^2)}} \right )
\label{Li-term}
\eqa
describes contributions involving ratios of internal and external masses.

Using (\ref{c0expansion}, \ref{bij-mass-ratios}, \ref{Li-term})
and  Appendix 2 and the results in \cite{Hag}, the implied
 asymptotic results for $C_i$ are
\bqa
&& C_{11} \simeq -  \frac{(\ln s_2)^2}{2 s}
+\frac{\ln s_{12} }{s}  -\frac{ L_{223}
+L_{121}-b_0^{(12)}+2 }{s}  ~~, \label{C11-asym} \\
&& C_{12}
 \simeq -\frac{\ln s_{23}}{s} + \frac{2-b_0^{(23)}}{s}    ~~, \label{C12-asym} \\
&& C_{24}
\simeq -  \frac{\ln s_\mu }{4}+\frac{\Delta +3 }{4} ~~, \label{C24-asym} \\
&& C_{21}
\simeq \frac{(\ln s_2 )^2}{2 s}-\frac{3 \ln s_{12}}{2 s}
+\frac{L_{223}+L_{121}-b_0^{(12)}+b_1^{(12)}+3}{s}  ~~, \label{C21-asym}\\
&& C_{23}
\simeq \frac{\ln s_{23}}{s} + \frac{2 b_0^{(23)}-5}{2s}   ~~, \label{C23-asym}\\
&& C_{22}
\simeq \frac{\ln s_{23}}{2 s} - \frac{1+b_1^{(23)}}{s}    ~~, \label{C22-asym}\\
&& C_{001}
\simeq \frac{\ln s_\mu }{6}-\frac{\Delta}{6}-\frac{19}{36}
~~, \label{C001-asym}\\
&& C_{002}
\simeq \frac{\ln s_\mu }{12}-\frac{\Delta}{12}-\frac{2}{9}
 ~~,\label{C002-asym} \\
&& C_{111}
\simeq -\frac{ (\ln s_2 )^2 }{2 s} + \frac{11 \ln s_{12} }{6 s}
 -\frac{ L_{223}+L_{121}}{s}+\frac{b_0^{(12)}-b_1^{(12)}+b_{21}^{(12)}}{s}
 -\frac{67}{18s} ~, \label{C111-asym}\\
&& C_{112}\simeq  -\frac{\ln s_{23} }{s}-\frac{b_0^{(23)}}{s}
 +\frac{17}{6s}      ~~,\label{C112-asym}\\
&& C_{122}
\simeq  -\frac{\ln s_{23}}{2 s}+ \frac{b_1^{(23)}}{s}+
\frac{7}{6s}     ~~, \label{C122-asym} \\
&& C_{222}
\simeq -\frac{\ln s_{23} }{3 s}-\frac{b_{21}^{(23)}}{s}  + \frac{13}{18s} ~~,
\label{C222-asym}
\eqa
where terms of  ${\cal O}(m_i^2/s)$ relative to those kept, have been neglected.
Among the PV functions listed above, we note
that $B_i$, $C_{24}$, $C_{001}$ and $C_{002}$), are the only
divergent ones.

\subsection{Asymptotic D-functions}

Based on Fig.\ref{box-fig} and the definition (\ref{Dkin1}),
we are interested in the asymptotic D functions in the
domain
\bq
(|s|,~|t|,~|u|) \gg (|p_j^2|,~ m_i^2 )~~,
\eq
\noindent
which means large energy and momentum transfer squares,
and fixed angles  different from 0 or $\pi$.
Defining also
\bqa
&& r_{ts} =\frac{-t-i\epsilon }{-s-i\epsilon}
~~~,
\nonumber \\
&&  t_1 =\frac{-t-i\epsilon}{m_1^2}~~~,
~~~ t_2=\frac{-t-i\epsilon}{m_2^2}~~~,
~~~t_3=\frac{-t-i\epsilon}{m_3^2}~~~,~~~ t_4=\frac{-t-i\epsilon}{m_4^2}~~ ,
\nonumber \\
&& s_1 =\frac{-s-i\epsilon}{m_1^2}~~~,
~~~ s_2=\frac{-s-i\epsilon}{m_2^2}~~~, ~~~ s_3=\frac{-s-i\epsilon}{m_3^2}
~~~, ~~~ s_4 =\frac{-s-i\epsilon}{m_4^2} ~~~,
\nonumber \\
&&s_{ij}=\frac{-s -i\epsilon}{m_im_j}~~~,
~~~t_{ij}=\frac{-t -i\epsilon}{m_im_j}~~~, \label{Dkin2}
\eqa
the basic expression of Denner and Roth \cite{Denner-Roth} is
\bqa
\label{d0expansion}
D_0  & \simeq & \frac{1}{s t}
\Bigg \{ -(\ln r_{ts} )^2-\pi^2+\frac{1}{2}
\left [ (\ln t_2 )^2+ (\ln t_4 )^2+ (\ln  s_3 )^2
+(\ln s_1 )^2\right ]\nonumber \\
&&
+L_{223} +L_{121}+ L_{441}+ L_{343}+ L_{334}
+L_{232}+ L_{112}+ L_{414} \Bigg \},
\eqa
where the neglected terms are suppressed by an additional factor of
either $s$ or $t$ or $u$. To the same accuracy, the results
of Appendix 2 imply
\bqa
D_{11} & \simeq  & \frac{(u-t ) (\ln r_{ts} )^2}{2 s t u}
-\frac{ (\ln s_3 )^2}{2 s t}
-\frac{ (\ln t_2 )^2}{2 s t}
-\frac{ (\ln t_4 ))^2}{2 s t}-\frac {\pi^2 (t-u )}{2 s t u}
\nonumber \\
&& -\frac{(  L_{121} +  L_{223} +  L_{232} + L_{334} +
L_{343} + L_{441} ) }{ s t } ~~ \label{D11-asym}\\
D_{12} & \simeq  & \frac{ (\ln r_{ts} )^2}{2 s t}
-\frac{ (\ln s_3 )^2}{2 s t} -\frac{ (\ln t_4 )^2}{2 s t}
-\frac{ L_{232}+ L_{334}+ L_{343}+ L_{441}}{s t}
+\frac{\pi^2}{2 s t}~~, \label{D12-asym}  \\
D_{13} & \simeq & -\frac{(\ln r_{ts})^2}{2 s u}-\frac{(\ln t_4 )^2}{2 s t}
-\frac{\pi^2}{2 s u}-\frac{L_{343}+L_{441}}{s t}~~, \label{D13-asym} \\
D_{27} & \simeq  & -\frac{ (\ln r_{ts} )^2}{4 u}-\frac{\pi^2}{4 u}~~,
\label{D27-asym} \\
D_{21} & \simeq  & -\frac{(t^2+u^2 ) (\ln r_{ts})^2}{2 s t u^2}
  +\frac{(\ln t_2 )^2 +(\ln t_4)^2+ (\ln s_3)^2}{2 s t}
 - \frac{\ln r_{ts}}{su}-\frac{\ln s_{12}+\ln t_{14}}{st}
\nonumber \\
&&  +\frac{L_{121} +L_{223}+ L_{232}+ L_{334}+ L_{343} +L_{441}
-b_0^{(12)}-b_0^{(14)}}{ s t}
\nonumber \\
&&   -\frac{\pi^2 (t^2+u^2 )}{2 s t u^2}+\frac{4}{st}~, \label{D21-asym} \\
D_{22} & \simeq & - \frac{\ln (s_{14}  t_{23} )}{ s t}
+\frac{(\ln s_3 )^2+(\ln t_4 )^2-(\ln r_{ts} )^2}{2 s t}
+\frac{8-\pi^2}{2 s t}
\nonumber \\
&& +\frac{L_{232}+ L_{334}+ L_{343}+ L_{441}-b_0^{(14)}-b_0^{(23)} }{ s t}
~~, \label{D22-asym} \\
D_{23}& \simeq  &  -\frac{t (\ln r_{ts})^2}{2 s u^2}
+\frac{ (\ln t_4 )^2}{2 s t}
 - \frac{\ln r_{ts}}{su}-\frac{\ln t_{14}+\ln t_{34}}{st}
 +\frac{8u^2-\pi^2 t^2}{2 s t u^2}
\nonumber \\
&& +\frac{  L_{343}+ L_{441}-b_0^{(14)}-b_0^{(34)} }{ s t}~~, \label{D23-asym} \\
D_{24}& \simeq  & +\frac{(\ln s_3 )^2+(\ln t_4 )^2-(\ln r_{ts} )^2}{2 s t}
-\frac{\ln s_{14}}{st}+\frac{4-\pi^2}{2 s t}
\nonumber \\
&& +\frac{ L_{232}+ L_{334}+ L_{343} +L_{441}-b_0^{(14)}}{ s t}~, \label{D24-asym} \\
D_{25}& \simeq  &   - \frac{\ln r_{ts}}{su}-\frac{\ln t_{14}}{st}
-\frac{t (\ln r_{ts} )^2}{2 s u^2}\nonumber \\
&&
+\frac{(\ln t_4 )^2}{2 s t}
 +\frac{L_{343} +L_{441}-b_0^{(14)}}{s t} +\frac{ 4 u^2-\pi^2 t^2}{2 s t u^2}
 ~~, \label{D25-asym} \\
D_{26} & \simeq  & +\frac{(\ln r_{ts} )^2}{2 s u}
+\frac{(\ln t_4 )^2}{2 s t}  - \frac{\ln  t_{14} }{ s t}
+\frac{L_{343}+ L_{441}-b_0^{(14)}}{s t}
+\frac{\pi^2 t+4u}{2 s t u }~, \label{D26-asym}\\
D_{001} & \simeq & \frac{(u-t ) (\ln  r_{ts} )^2}{8 u^2}
-\frac{\ln r_{ts}}{4 u}-\frac{\pi^2 (t-u)}{8 u^2}~, \label{D001-asym}\\
D_{002} & \simeq  & \frac{(\ln r_{ts} )^2}{8 u} +\frac{\pi^2}{8 u}
~, \label{D002-asym}\\
D_{003} & \simeq  & - \frac{t (\ln r_{ts} )^2}{8 u^2}
-\frac{\ln r_{ts} }{4 u}-\frac{\pi^2 t}{8 u^2}~, \label{D003-asym} \\
D_{111} & \simeq  & -\frac{ (2 t- u)\ln r_{ts}}{2 s u^2}
+\frac{3}{2st}(\ln s_{12}+\ln t_{14})
+\frac{(u^3-t^3 ) (\ln r_{ts} )^2}{2 s t u^3}
\nonumber \\
&& -\frac{(\ln s_3 )^2 +(\ln t_2 )^2 +(\ln t_4 )^2}{2 s t}
 -\frac{L_{121} + L_{223}+ L_{232}+ L_{334}+ L_{343} + L_{441}}{s t}
\nonumber \\
&& +\frac{b_0^{(12)}+b_0^{(14)}-b_1^{(12)}- b_1^{(14)}}{st}
-\frac{(t+11u ) u^2+\pi^2 (t^3-u^3 )}{2 s t u^3}~, \label{D111-asym}\\
D_{112}& \simeq & \frac{3 \ln s_{14} }{2 s t}
+\frac{ (\ln r_{ts})^2 - (\ln s_3 )^2 - (\ln t_4 )^2}{2 s t}
 -\frac{L_{232}+ L_{334}+ L_{343}+ L_{441}}{st}
 \nonumber \\
&& +\frac{ b_0^{(14)}-b_1^{(14)}}{st}+\frac {\pi^2-5}{2 s t}~~, \label{D112-asym} \\
D_{113}& \simeq  &  -\frac{(2t-u ) \ln r_{ts}  }{2 s u^2}
+\frac{3  \ln  t_{14} }{2  st}
-\frac{t^2  (\ln r_{ts} )^2}{2 s u^3}
-\frac{ (\ln t_4 )^2}{2 s t}
 -\frac{ L_{343}+ L_{441}}{s t}
\nonumber \\
&& +\frac{ b_0^{(14)}-b_1^{(14)}}{st}-\frac{\pi^2 t^3+t u^2+6 u^3}{2 st u^3}
~, \label{D113-asym} \\
D_{122} & \simeq  & \frac{3\ln  s_{14}+2 \ln  t_{23}   }{2 s t}
+\frac{ (\ln r_{ts}  )^2 - (\ln s_3 )^2 - (\ln t_4 )^2}{2 s t}
-\frac{ L_{232}+ L_{334}+ L_{343}+ L_{441}}{ s t}
\nonumber \\
&& + \frac{b_0^{(14)}+b_0^{(23)}-b_1^{(14)}}{st}
+\frac{\pi^2-10}{2 s t}~~, \label{D122-asym}\\
D_{222} & \simeq  & \frac{3 (\ln  s_{14}+\ln  t_{23}) }{2 s t}
+\frac{( \ln r_{ts} )^2 - (\ln s_3 )^2 - ( \ln t_4 )^2}{2 s t}
-\frac{ L_{232} + L_{334} + L_{343} + L_{441}}{s t}
\nonumber \\
&& +\frac{b_0^{(14)}+b_0^{(23)}-b_1^{(14)}-b_1^{(23)}}{st}
+\frac{\pi^2-12}{2 s t}~~, \label{D222-asym}\\
D_{223} & \simeq & \frac{3 \ln  t_{14} }{2 s t}
-\frac{ (\ln r_{ts} )^2}{2 s u}-\frac{ (\ln t_4 )^2}{2 s t}
-\frac{L_{343}+ L_{441}}{s t}+\frac{b_0^{(14)}-b_1^{(14)}}{s t}
-\frac{\pi^2t+6u}{2 s t u}~~, \label{D223-asym}\\
D_{123} & \simeq &  \frac{\ln r_{ts}  }{2 s u}
+\frac{3 \ln  t_{14} }{2 s t }
+\frac{(t-u ) (\ln r_{ts} )^2}{4 s  u^2}
-\frac{ ( \ln t_4 )^2}{2 s t}  -\frac{ L_{343} + L_{441}}{s t}
\nonumber \\
&& +\frac{b_0^{(14)}-b_1^{(14)}}{st}
+\frac{\pi^2 t(t-u)-12 u^2 }{4 s t u^2}~~, \label{D123-asym} \\
D_{133}& \simeq  & -\frac{(2 t-u ) \ln  r_{ts}  }{2 s u^2}
+\frac{(3\ln t_{14}+2\ln t_{34} }{2 s t }
-\frac{ (\ln t_4 )^2}{2 s t}
-\frac{t^2 (\ln r_{ts} )^2}{2 s u^3}
\nonumber \\
&& - \frac{ L_{343}+ L_{441}}{s t}
+\frac{b_0^{(14)}+b_0^{(34)}-b_1^{(14)}}{st}
-\frac{\pi^2 t^3+t u^2+11 u^3}{2 s t u^3}~~, \label{D133-asym}\\
D_{233} & \simeq  &  \frac{\ln r_{ts}  }{s u}
+\frac{(3\ln  t_{14}+2\ln  t_{34}) }{2 s t }
+\frac{t  (\ln r_{ts} )^2}{2 s u^2}
 -\frac{ (\ln t_4 )^2}{2 s t}
-\frac{ L_{343} + L_{441}}{s t}
\nonumber \\
&& +\frac{b_0^{(14)}+b_0^{(34)}-b_1^{(14)}}{st}
 +\frac{\pi^2 t^2-11 u^2 }{2 s t u^2} ~~, \label{D233-asym} \\
D_{333}& \simeq  & -\frac{ (2 t-u ) \ln r_{ts} }{2 s u^2}
+\frac{3( \ln  t_{14}+\ln  t_{34}) }{2 s t }
-\frac{t^2 (\ln r_{ts} )^2}{2 s u^3}-\frac{(\ln t_4 )^2}{2 s t}
\nonumber \\
&&  - \frac{ L_{343}+ L_{441}}{s t}
+\frac{b_0^{(14)}+b_0^{(34)}-b_1^{(14)}-b_1^{(34)}}{st}
-\frac{\pi^2 t^3+t u^2+ 14 u^3}{2 s t u^3}~~. \label{D333-asym}
\eqa\\

\section{Discussion of the results}

The main results of this paper are,
apart from the simple case of (\ref{B-asym})
for the $B$ functions, (\ref{C11-asym}-\ref{C222-asym})
for the $C$ functions and (\ref{D11-asym}-\ref{D333-asym}) for the $D$ functions.
In these analytic expressions one recognizes:
\medskip

\begin{enumerate}
\item The true leading quadratic logarithms
which in SM or SUSY only arise from gauge boson exchanges,
and the  linear logarithmic terms  arising  also  from gauge boson exchanges,
as well as from many other exchanges; see \cite{SMrules, MSSMrules}.
Note also that terms like
\bq
\ln^2\frac{-s-i\epsilon}{m^2}~~~,\qquad
\ln^2\frac{-t-i\epsilon}{m^2}~~~,\qquad
\ln^2\frac{-u-i\epsilon}{m^2}~~~,
\eq
generate not only $\ln^2s$ contributions, but also subleading,
angular dependent and true constant terms, as seen in
\bqa
\ln^2{-s-i\epsilon\over m^2} &=& \left(\ln{s\over m^2}-i\pi\right)^2=
\ln^2{s\over m^2}-2i\pi \ln{s\over m^2}-\pi^2, \nonumber \\
\ln^2{-t-i\epsilon\over m^2} &=& \ln^2\left|{t\over m^2}\right|=
\ln^2s+2\ln s\ln{1-\cos\theta\over2}
+\ln^2{1-\cos\theta\over2}+{\cal O}\left({m^2\over s}\right), \nonumber \\
\ln^2{-u-i\epsilon\over m^2} &=& \ln^2\left|{u\over m^2}\right|=
\ln^2s+2\ln s\ln{1+\cos\theta\over2}
+\ln^2{1+\cos\theta\over2}+{\cal O}\left({m^2\over s}\right). \nonumber
\eqa
Linear logarithms also appear as
\bqa
\ln{-s-i\epsilon\over m^2} &=& \ln{s\over m^2}-i\pi, \nonumber \\
\ln{-t-i\epsilon\over m^2} &=& \ln{s\over m^2}+\ln{1-\cos\theta\over2}
+{\cal O}\left({m^2\over s}\right), \nonumber \\
\ln{-u-i\epsilon\over m^2} &=& \ln{s\over m^2}+\ln{1+\cos\theta\over2}
+{\cal O}\left({m^2\over s}\right), \nonumber
\eqa
in which mass suppressed terms have not been written
explicitly.\\

\item The {\em constant} terms which consist,
as one sees explicitly in Sect.(3.1, 3.2, 3.3),
of  true constant numbers (see for instance (\ref{C11-asym})), as well as
of logarithmic or $\mbox{Li}_2$ functions
involving  ratios of masses or  other kinematical quantities, as they
appear in $L_{ijk}$ and $b_i$.
They are called {\em constant} because they are indeed
$s$-independent, but in some cases
they may contain angular dependencies. A priori these
$L_{ijk}$ and $b_i$ quantities contain  all internal and
external masses and mixings.
\end{enumerate}

The omitted terms in all our asymptotic expression
are  mass-suppressed like $m^2/s$, relative to the retained ones
and control the approach to asymptopia. In the remaining  figures we illustrate
this approach with a few examples, showing how the
asymptotic PV functions match with the exact ones at
high energies. This  provides a useful insight about the properties
of the various PV functions.

\begin{figure}
\centering
\epsfig{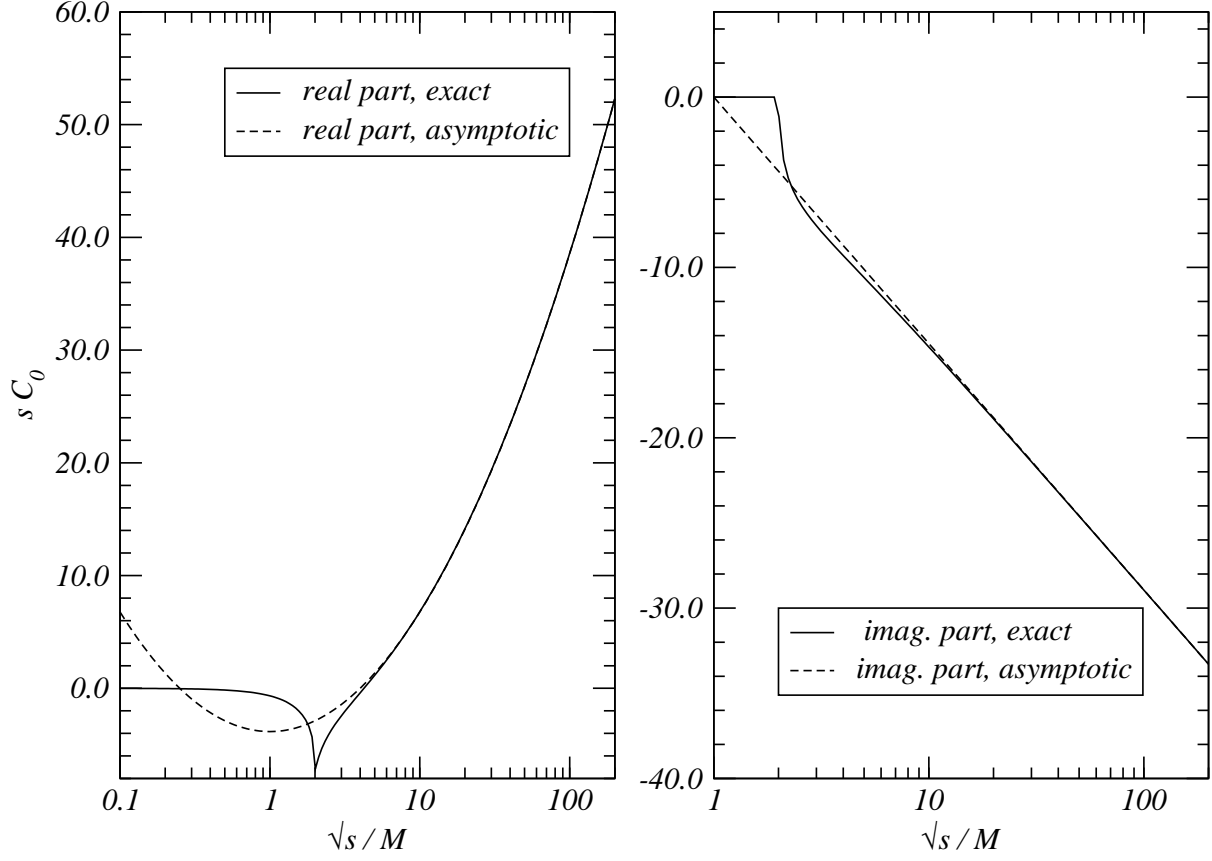}
\vspace{1.5cm}
\caption{Comparison of the exact  and asymptotic results for
$sC_0$, as a function of $\sqrt{p_3^2}/M \equiv \sqrt{s}/M$, ~
at ~ $m_i^2 = p_1^2=p_2^2 = M^2$. Real and Imaginary parts are studied in the
left and right panels respectively.}
\label{fig:c0}
\end{figure}

We begin with the basic  $C_0$ function
where, for illustration, we consider the simplest possible
kinematical configuration with all internal masses put at a common scale,
and the external squared momenta of two legs set also at a common mass scale; \ie
\bq
m_i^2 =m^2~~,~~ p_1^2=p_2^2 = M^2 ~~ ,~~ p_3^2=s ~~. \label{C-point}
\eq
This way, the deviations between  the exact and asymptotic results can be
studied as a function of the   dimensionless
parameter $\sqrt{s}/M$. In Fig.\ref{fig:c0} this is done
for the real and imaginary parts of the dimensionless quantity $s C_0$,
choosing also $m=M$. As seen there, $s C_0$ becomes predominantly real
at asymptotic $s$, and
the approximate expression (\ref{c0expansion}) is quite accurate for
 $\sqrt{s}/M\gsim 5$.

A similar analysis  is done for  $s^2 D_0$ in Fig.\ref{fig:d0}, where
a common scale is again chosen as
\bq
m_i^2 =m^2~~,~~ p_j^2= M^2 ~~ ,~~ t=-\frac{s}{2}~~ , \label{D-point}
\eq
and  (\ref{Dkin1}) is used.
As seen from  Fig.\ref{fig:d0}, (where we have again for simplicity chosen
$m=M$), $s^2 D_0$ become predominantly imaginary at asymptotic $s$,
and the exact and asymptotic results almost coincide
for $\sqrt{s}/M \gtrsim 5$.  \\

Another  interesting application of the  results  in Sect.3,  concerns
 combinations of PV functions in which  the asymptotic   logarithmic contributions
  cancel out,  and only  mass-independent constants remain
asymptotically. Examples of such combinations are
\bq
\begin{array}{l}
s(C_{23}+C_{12})~~,~~  s  D_{27}~~,~~ \\
 s D_{00i}~~,~~ s^2(D_{112}-D_{123}+D_{24}-D_{26})~~,\\
s^2(D_{113}-D_{112}+D_{122}-D_{123}+D_{25}-D_{24}+D_{22}-D_{26})~~,
\end{array}
\eq
which  often appear in some SUSY applications \cite{Wpap}.

\begin{figure}
\centering
\epsfig{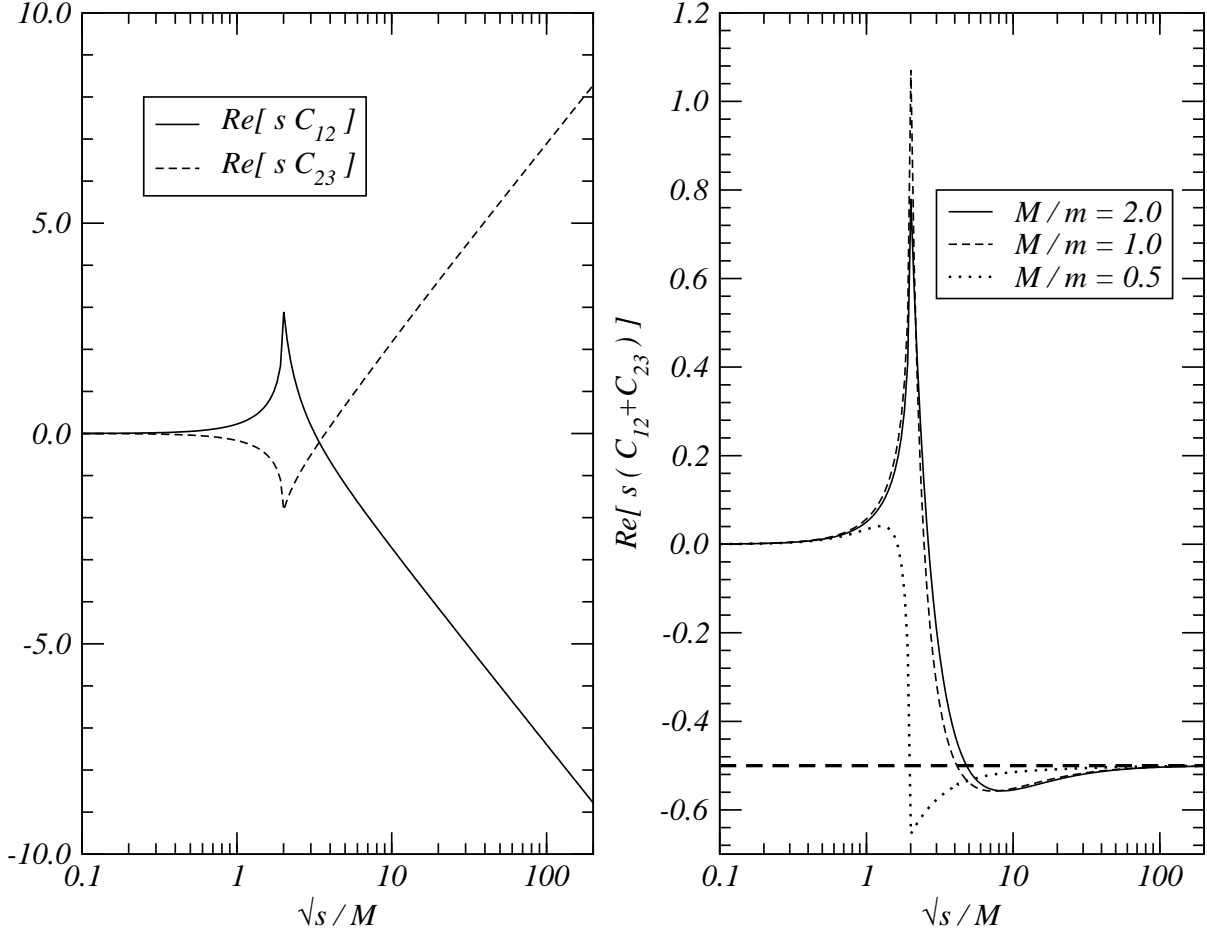}
\vspace{1.5cm}
\caption{ The left panel shows the exact results for $Re[sC_{12}]$
and $Re[sC_{23}]$,
as functions of $\sqrt{s}/M$, with the other parameters chosen as in
(\ref{C-point}). It clearly indicates the asymptotic logarithmic behavior.
As seen in the right panel though, the logarithmic contribution
cancels out for high $\sqrt{s}/M$ in the combination $Re[s(C_{12}+C_{23})]$,
and only a universal constant remains.
The asymptotic  predictions from (\ref{C12-asym}, \ref{C23-asym}),
are described  by the horizontal dashed line
which agrees with the exact results at high $\sqrt{s}/M$ .}
\label{fig:split}
\end{figure}

As a first example, we plot in the right panel of Fig.\ref{fig:split}
the real part of  $s(C_{23}+C_{12})$,  as a function of $\sqrt{s}/M$;
the other parameters chosen as in (\ref{C-point}), while allowing
for  three ratios $M/m = 2, 1, 1/2$.
As seen in the left panel of the same  figure,  the logarithmic terms
strongly  dominate the  exact results for  $Re[s C_{23}]$ and $Re[sC_{12}]$
at high $s$. But as the  right panel indicates, these logarithmic contributions
cancel out in  $Re[s(C_{23}+C_{12})]$,
and only a tiny constant contribution remains
at  high $s$, which seems independent of the mass ratio $M/m$.
The prediction from our asymptotic expansion
is also shown as an horizontal dash line, which agrees with the exact result
for  $\sqrt{s}/M \gtrsim 5$.

\begin{figure}[p]
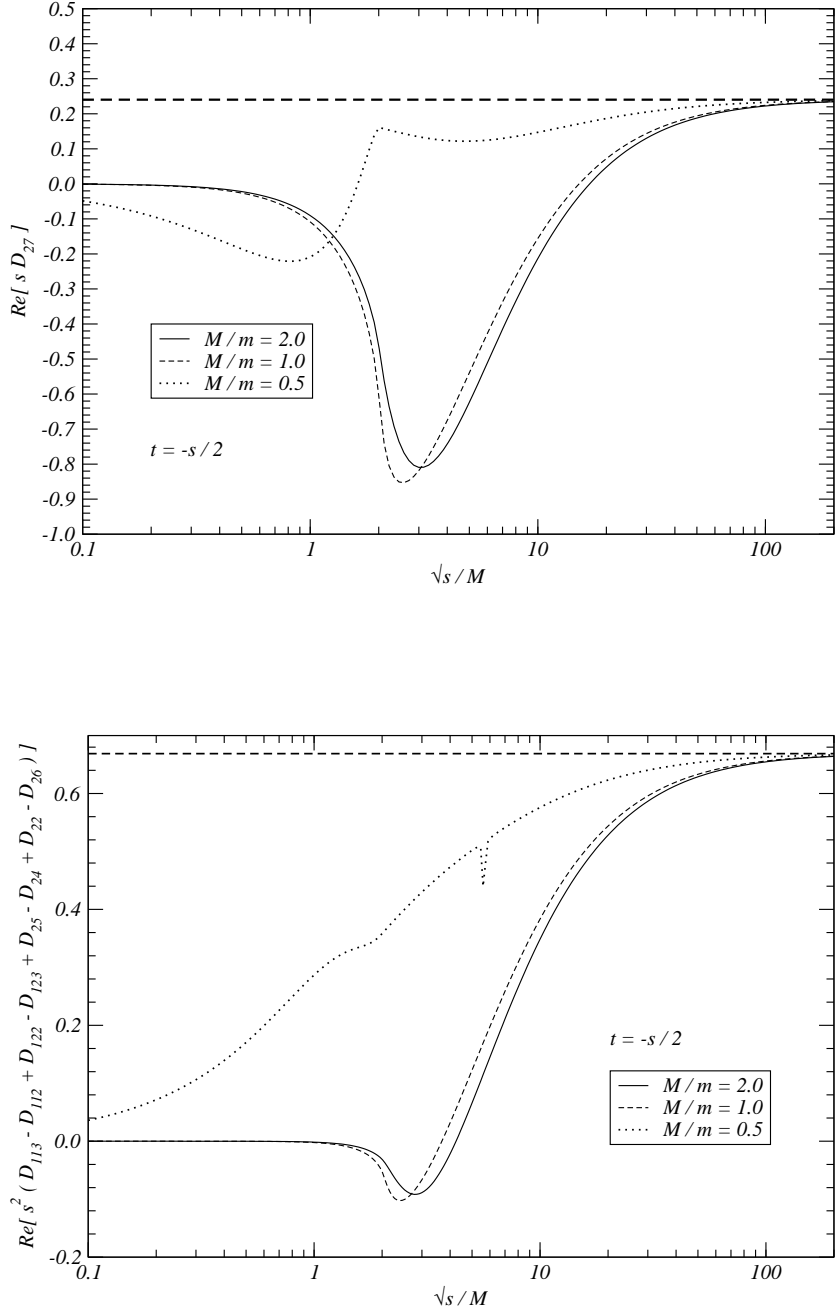

\vspace{-1.cm}
\[
\epsfig{file=D27.eps, width=11.cm, angle=0}
\]
\vspace{1.cm}
\[
\epsfig{file=Z2.eps, width=11.cm, angle=0}
\]
\caption{In the upper panel,  $Re[s D_{27}]$, which has no
asymptotic logarithmic contribution, is plotted against $\sqrt{s}/M$,
with the remaining parameters fixed as in (\ref{D-point}).
The exact results for  $Re[s D_{27}]$  behave like
a mass independent constant at high  $\sqrt{s}/M$. This constant
agrees with the asymptotic predictions from Sect. 3,
described  by the horizontal dashed line. In the lower channel a
similar analysis is done for the combination
$s^2\,(D_{113}-D_{112}+D_{122}-D_{123}+D_{25}-D_{24}+D_{22}-D_{26})$,
which has similar mathematical properties. }
\label{fig:d27}
\end{figure}

As a second example we present in Fig.\ref{fig:d27} a similar analysis for
$Re[sD_{27}]$ (upper panel) and
 $s^2(D_{113}-D_{112}+D_{122}-D_{123}+D_{25}-D_{24}+D_{22}-D_{26})$ (lower panel),
 plotted as  functions of $\sqrt{s}/M$, with
the other parameters chosen as in (\ref{D-point}), using again
$M/m = 2, 1, 1/2$. In both cases the asymptotic predictions from the results in Sect.
3 are indicated by the horizontal dash lines.\\

\begin{figure}
\centering
\epsfig{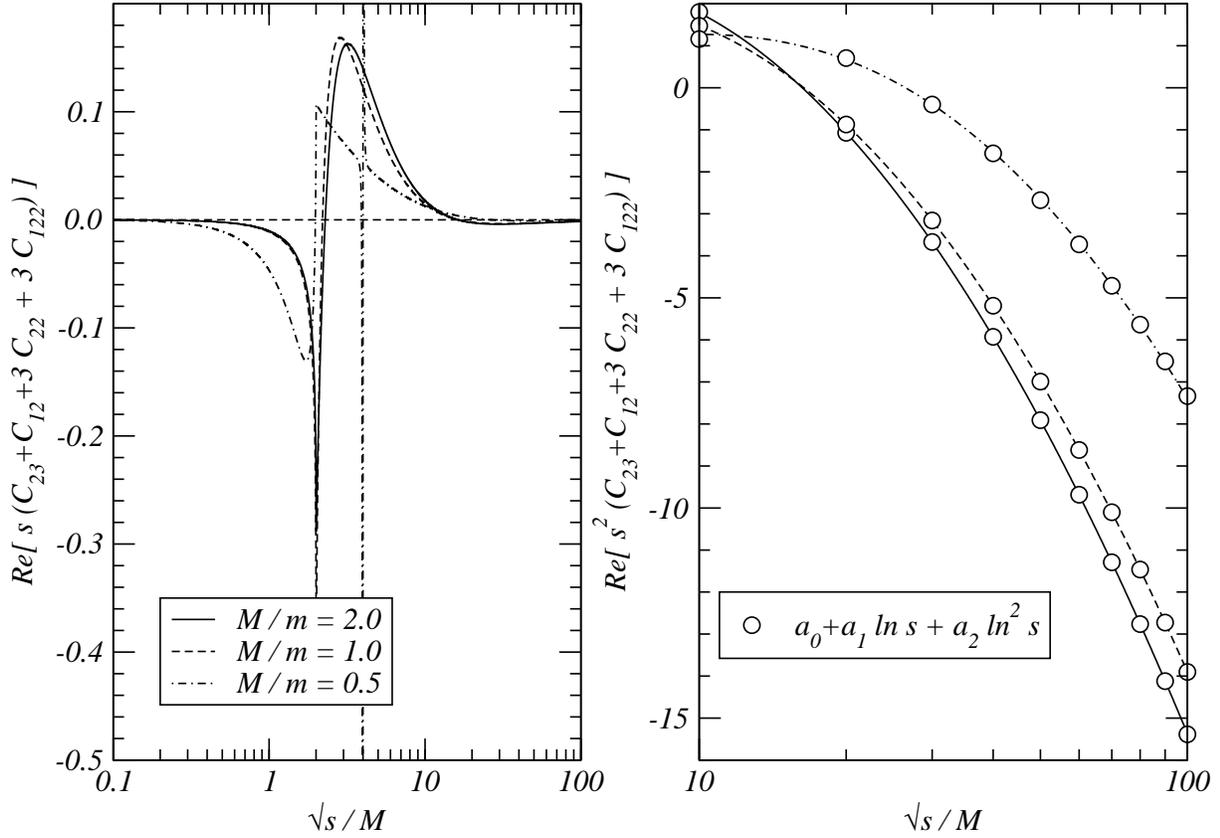}
\vspace{1.5cm}
\caption{The left panel presents the asymptotically vanishing combination
$s(C_{23}+C_{12}+3C_{22}+3C_{122})$, as a function of $\sqrt{s}/M$,
 with the remaining parameters chosen as in \ref{C-point}. The asymptotic
 vanishing is evident. The model dependence of the approach
 to this zero-value is shown in the right
 panel obtained by multiplying the whole expression
 by an  additional $s$ factor, which leads to a quantity behaving like
 a  quadratic polynomial in $\ln s$.}
\label{fig:w1}
\end{figure}

We now turn to  PV combinations
in which   the constant terms also
cancel out asymptotically, together with the logarithms.
In such a case,  only model
dependent terms, suppressed by an extra
power of $s$, remain at high energies.
An example of such combinations is  given by
\bq
s (C_{23}+C_{12}+3C_{22}+3C_{122})~~, \label{no-constant}
\eq
which is plotted  in the left panel of  Fig.\ref{fig:w1},
as a function of  $\sqrt{s}/M$, choosing again the internal and
external masses as in (\ref{C-point}).
Its  asymptotic vanishing is evident.
The right panel of Fig.\ref{fig:w1}
shows what is obtained when (\ref{no-constant}) is multiplied by
 an additional $s$ factor,
which allows to inspect its model dependence  at high  $\sqrt{s}/M$,
where it can at most increase like a power of a logarithm.
As the open circles show, a quadratic polynomial in $\ln s$
is accurately describing this rise,
in the present case.  Similar results are obtained for other such combinations.\\

\section{Conclusions and Outlook}

The above asymptotic expressions of $B$, $C$, $D$ functions
should be useful for the analysis of many SM and
MSSM (or NMSSM) 2-to-2 body processes at LHC and future
high energy colliders. This is particularly true in situations where the energies may be
much higher than all internal and external masses, and the scattering angles
are kept fixed.

Particularly  for MSSM (or NMSSM), the above expressions
may be useful for exploiting  the intriguing HC property,
which induces logarithmically increasing 1-loop contributions
to the total helicity conserving amplitudes;
while striking cancellations appear for the amplitudes violating HC.

Below we illustrate this assertion with the case of the
process $ug\to dW$, for which the analysis of  \cite{Wpap} has
revealed peculiar virtual SUSY effects in the helicity
amplitudes. Particularly for the helicity violating  amplitudes,
 spectacular high energy cancellations
have been found. For these amplitudes, the one loop electroweak
corrections have no leading logarithms, but they tend instead to
a  {\em constant} limit in SM, which in MSSM (or NMSSM) exactly vanishes,
due to an opposite SUSY  contribution.
Thus, in MSSM, all  helicity violating amplitudes are of order
${\cal O}(m^2/s)$ and possibly negligible  at LHC  energies.

Since the general proof of the HC
theorem in \cite{heli} neglected electroweak breaking,
it is important to check in various cases how possible  constant terms
involving ratios of masses combine to assure the validity od the theorem.
The above $B$, $C$, $D$
expressions should be useful for this.
An example of this is seen in \cite{Wpap}.

Beyond this though, the application in \cite{Wpap} also indicates that,
 although HC is an
asymptotic theorem, it may be important  at the LHC range,
 where it  may strongly reduce the number of  important amplitudes
 to just those respecting HC; thereby
 simplifying the analysis.

For this reason, the above high energy
approximations of the $B$, $C$, $D$ functions should
allow to make \underline{quantitative}
predictions for the physical amplitudes.
This must be a valuable improvement
with respect to the leading logarithmic level in two aspects.
Firstly, the leading logarithms only test the gauge and
Yukawa structures. Although this is an   important
step in SUSY checking, the {\em constant} terms should  open
the door to deeper tests of the SUSY structure.
Second, the comparison with experimental results
of the LL  approximation,
requires delicate experimental  fits of logarithmic expressions,
which are only realizable if several points in the high energy range
are available.
On the opposite, the more complete asymptotic
expressions written in this paper
are directly usable at any given high energy point.
These expressions are analytically simple and
can be easily put in a code allowing quick computations
for any MSSM benchmark.

\newpage

\appendix

\renewcommand{\thesection}{Appendix \arabic{section}}
\renewcommand{\theequation}{A.\arabic{equation}}
\setcounter{equation}{0}

\section{Relation with the LoopTools conventions}

In the LoopTools library \cite{looptools} the following
momenta are defined
\bqa
k_1 &=& p_1 ~, \nonumber\\
k_2 &=& p_1+p_2  ~, \nonumber\\
k_3 &=& p_1+p_2+p_3 ~, \nonumber
\eqa
leading to  the  tensorial decomposition
\bqa
B^\mu &=& k_1^\mu B_1^L ~, \nonumber\\
B^{\mu\nu} &=& k_1^\mu k_1^\nu B_{11}^L+g^{\mu\nu} B_{00}^L ~, \nonumber
\eqa
\bqa
C^\mu &=& k_1^\mu C_1^L+k_2^\mu C_2^L ~, \nonumber\\
C^{\mu\nu} &=& \sum_{ij=1}^2 k_i^\mu k_j^\nu C_{ij}^L+
g^{\mu\nu} C_{00}^L ~, \nonumber\\
C^{\mu\nu\rho} &=& \sum_{i,j,l=1}^2 k_i^\mu k_j^\nu k_l^\rho C_{ijl}^L +
\sum_{i=1}^2(g^{\mu\nu}k_i^\rho + g^{\mu\rho}k_i^\nu + g^{\nu\rho}k_i^\mu)
 C_{00i}^L ~, \nonumber
\eqa
\bqa
D^\mu &=& k_1^\mu D_1^L + k_2^\mu D_2^L + k_3^\mu D_3^L ~, \nonumber\\
D^{\mu\nu} &=& \sum_{i,j=1}^3 k_i^\mu k_j^\nu D_{ij}^L +
g^{\mu\nu}~  D_{00}^L \nonumber
\eqa
where the superscript $L$ denotes the PV functions in the
LoopTools notation.
Comparing with (\ref{B-tensor}, \ref{C-tensor}, \ref{D-tensor}),
their relations with the PV functions in the Hagiwara decomposition is
\bqa
B_1 &=& B_1^L ~~, \nonumber\\
B_{21} &=& B_{11}^L ~~, \nonumber\\
B_{22} &=& B_{00}^L ~~,\\[0.5cm]
C_{11} &=& C_1^L + C_2^L ~~, \nonumber\\
C_{12} &=& C_2^L ~~,\nonumber\\
C_{21} &=& C_{11}^L+2C_{12}^L+C_{22}^L ~~, \nonumber\\
C_{22} &=& C_{22}^L ~~, \nonumber\\
C_{23} &=& C_{12}^L + C_{22}^L ~~, \nonumber\\
C_{24} &=& C_{00}^L \nonumber\\
C_{001} &=& C_{001}^L+C_{002}^L ~~, \nonumber\\
C_{002} &=& C_{002}^L ~~, \nonumber\\
C_{111} &=& C_{111}^L+3C_{112}^L+3C_{122}^L+C_{222}^L ~~, \nonumber\\
C_{222} &=& C_{222}^L ~~, \nonumber\\
C_{112} &=& C_{112}^L+2C_{122}^L+C_{222}^L ~~,\nonumber\\
C_{122} &=& C_{122}^L+C_{222}^L  ~~,
\eqa
\bqa
D_{11} &=& D_1^L+D_2^L+D_3^L  ~~, \nonumber\\
D_{12} &=& D_2^L+D_3^L  ~~, \nonumber\\
D_{13} &=& D_3^L  ~~, \nonumber \\
D_{21} &=& D_{11}^L+D_{22}^L+D_{33}^L+2(D_{12}^L+D_{13}^L+D_{23}^L)  ~~, \nonumber\\
D_{22} &=& D_{22}^L+2D_{23}^L+D_{33}^L  ~~, \nonumber\\
D_{23} &=& D_{33}^L  ~~, \nonumber\\
D_{24} &=& D_{12}^L+D_{13}^L+D_{22}^L+2D_{23}^L+D_{33}^L  ~~, \nonumber\\
D_{25} &=& D_{13}^L+D_{23}^L+D_{33}^L  ~~, \nonumber\\
D_{26} &=& D_{23}^L+D_{33}^L  ~~, \nonumber\\
D_{27} &=& D_{00}^L  ~~, \nonumber\\
D_{001} &=& D^L_{001}+D^L_{002}+D^L_{003} ~~, \nonumber\\
D_{002} &=& D^L_{002}+D^L_{003}  ~~, \nonumber\\
D_{003} &=& D^L_{003}  ~~, \nonumber\\
D_{111} &=& D^L_{111} +3 D^L_{112}+3 D^L_{113}+ 3 D^L_{122} +
6 D^L_{123}+3 D^L_{133}+ D^L_{222} + 3 D^L_{223} + 3 D^L_{233} + D^L_{333} ~~, \nonumber\\
D_{112} &=& D^L_{112} +D^L_{113}+2 D^L_{122}+ 4 D^L_{123}+ 2 D^L_{133}+ D^L_{222}
+ 3 D^L_{223} + 3 D^L_{233} + D^L_{333}  ~~, \nonumber\\
D_{113} &=& D^L_{113} + 2 D^L_{123} + 2 D^L_{133} + D^L_{223}
+ 2 D^L_{233} + D^L_{333}  ~~, \nonumber\\
D_{122} &=& D^L_{122} + 2 D^L_{123} + D^L_{133} + D^L_{222} + 3 D^L_{223} + 3 D^L_{233} + D^L_{333}\nonumber\\
D_{133} &=& D^L_{133} + D^L_{233} + D^L_{333} ~~, \nonumber\\
D_{123} &=& D^L_{123} + D^L_{133} + D^L_{223} + 2 D^L_{233} + D^L_{333} ~~,\nonumber\\
D_{222} &=& D^L_{222} + 3 D^L_{223} + 3 D^L_{233} + D^L_{333} ~~, \nonumber\\
D_{223} &=& D^L_{223} + 2 D^L_{233} + D^L_{333} ~~, \nonumber\\
D_{233} &=& D^L_{233} + D^L_{333} ~~, \nonumber\\
D_{333} &=& D^L_{333} ~~.
\eqa\\

\section{Reduction formalism for $C_{ijk}, ~D_{ijk}$.}

The following relations have not been explicitly written
in Hagiwara appendix. We write them below for completeness.\\

\subsection{$C_{ijk}$  formulae}
Same notation as in Hagiwara \cite{Hag}, with $(f_1,f_2)$, $X$ and
$B_i^{(jk)}$  taken respectively from  (D.32),  (D.31) and  (D.33) of \cite{Hag}.
\bqa
 X &= &\left ( \matrix{ 2p_1^2, 2p_1p_2 \cr
                   2 p_1 p_2, 2 p_2^2 }
                   \right )~~, \\[.3cm]
 \left (\matrix{C_{001} \cr C_{002} } \right ) &= &X^{-1}\cdot
\left (\matrix{B_{22}^{(13)}-B_{22}^{(23)}+f_1 C_{24} \cr
              B_{22}^{(12)}-B_{22}^{(13)}+f_2 C_{24} } \right ) ~~,  \\[.3cm]
\left (\matrix{C_{111} \cr C_{112} } \right ) &= &X^{-1}\cdot
\left (\matrix{B_{21}^{(13)}-B_{0}^{(23)}+f_1 C_{21}-4C_{001} \cr
              B_{21}^{(12)}-B_{21}^{(13)}+f_2 C_{21} } \right )~~,  \\[.3cm]
\left (\matrix{C_{122} \cr C_{222} } \right ) &= &X^{-1}\cdot
\left (\matrix {B_{21}^{(13)}-B_{21}^{(23)}+f_1 C_{22}  \cr
              -B_{21}^{(13)} +f_2 C_{22}-4 C_{002} } \right ) ~~,  \\[.3cm]
              \left (\matrix{C_{112} \cr C_{122} } \right ) &= &X^{-1}\cdot
\left (\matrix {B_{21}^{(13)} + B_{1}^{(23)}+f_1 C_{23}-2C_{002}  \cr
              -B_{21}^{(13)} +f_2 C_{23}-2 C_{001} } \right ) ~~.
\eqa\\

\subsection{$D_{ijk}$  formulae}
Expressions for $(f_1,f_2, f_3)$ and $X$ are taken respectively  from
  (D.37)  and (D.36) of \cite{Hag}. Using these we write
\bqa
D_{001}&=& \frac{1}{2} m_1^2 D_{11}-\frac{1}{4}
\left [ f_1 D_{21}+f_2 D_{24}+f_3 D_{25} +C_0^{(234)} \right ]~~, \nonumber\\
D_{002}&=& \frac{1}{2} m_1^2 D_{12}-\frac{1}{4}
\left [ f_1 D_{24}+f_2 D_{22}+f_3 D_{26} -C_{11}^{(234)} \right ]~~, \nonumber\\
D_{003}&=& \frac{1}{2} m_1^2 D_{13}-\frac{1}{4}
\left [ f_1 D_{25}+f_2 D_{26}+f_3 D_{23} -C_{12}^{(234)} \right ]~~,
\eqa
and
\bqa
R_{40}&= & f_1 D_{21}-C_0^{(234)}+C_{21}^{(134)}-4 D_{001} ~~, \nonumber \\
R_{41}&= & f_2 D_{21}-C_{21}^{(134)}+C_{21}^{(124)} ~~, \nonumber \\
R_{42}&= & f_3 D_{21}-C_{21}^{(124)}+C_{21}^{(123)} ~~, \nonumber \\
R_{44}&= & f_1 D_{24}+C_{21}^{(134)}+C_{11}^{(234)}-2 D_{002} ~~, \nonumber \\
R_{50}&= & f_1 D_{22}-C_{21}^{(234)}+C_{21}^{(134)} ~~, \nonumber \\
R_{56}&= & f_1 D_{23}-C_{22}^{(234)}+C_{22}^{(134)} ~~, \nonumber \\
R_{45}&= & f_2 D_{24}-C_{21}^{(134)}+C_{23}^{(124)}-2 D_{001} ~~, \nonumber \\
R_{51}&= & f_2 D_{22}-C_{21}^{(134)}+C_{22}^{(124)}-4 D_{002} ~~, \nonumber \\
R_{57}&= & f_2 D_{23}-C_{22}^{(134)}+C_{22}^{(124)} ~~, \nonumber \\
R_{46}&= & f_3 D_{24}-C_{23}^{(124)}+C_{23}^{(123)} ~~, \nonumber \\
R_{52}&= & f_3 D_{22}-C_{22}^{(124)}+C_{22}^{(123)} ~~, \nonumber \\
R_{58}&= & f_3 D_{23}-C_{22}^{(124)}-4 D_{003}  ~~, \nonumber \\
\eqa
\bqa
 X_3 &= &\left ( \matrix{ 2p_1^2, 2p_1p_2, 2 p_1 p_3 \cr
                   2 p_1 p_2, 2 p_2^2 , 2 p_2 p_3 \cr
                  2 p_1 p_3, 2 p_2 p_3, 2 p_3^2 }
                   \right ) ~~, \\[.3cm]
 \left ( \matrix{ D_{111} \cr D_{112} \cr D_{113} }\right )
 &=& X_3^{-1} \cdot \left ( \matrix{ R_{40} \cr R_{41} \cr R_{42} } \right )
 ~~, \\[.3cm]
\left ( \matrix{ D_{122} \cr D_{222} \cr D_{223} }\right )
 &=& X_3^{-1} \cdot \left ( \matrix{ R_{50} \cr R_{51} \cr R_{52} } \right )
 ~~, \\[.3cm]
 \left ( \matrix{ D_{133} \cr D_{233} \cr D_{333} }\right )
 &=& X_3^{-1} \cdot \left ( \matrix{ R_{56} \cr R_{57} \cr R_{58} } \right )
 ~~, \\[.3cm]
 \left ( \matrix{ D_{112} \cr D_{122} \cr D_{123} }\right )
 &=& X_3^{-1} \cdot \left ( \matrix{ R_{44} \cr R_{45} \cr R_{46} } \right )~~.
\eqa

In them we need addition
\bqa
C_{24}^{(123)} &=& \frac{1}{4}+\frac{1}{4} B_0^{(23)}+\frac{m_1^2}{2} C_0^{(123)}
-\frac{f_1}{4} C_{11}^{(123)}-\frac{f_2}{4} C_{12}^{(123)} ~~, \nonumber \\[.3cm]
\left ( \matrix{ C_{21}^{(123)} \cr C_{23}^{(123)} }\right )
 &=& \left ( \matrix{ 2 p_1^2 , ~2 p_1 p_2  \cr 2 p_1 p_2, ~ 2 p_2^2 } \right )^{-1}
  \left ( \matrix { B_1^{(13)}+B_0^{(23)}+f_1 C_{11}^{(123)}-2C_{24}^{(123)} \cr
 B_1^{(12)}-B_1^{(13)}+f_2 C_{11}^{(123)}  } \right )
\nonumber ~~, \\[.3cm]
 \left ( \matrix{ C_{23}^{(123)} \cr C_{22}^{(123)} }\right )
 &=& \left ( \matrix{ 2 p_1^2 , ~2 p_1 p_2  \cr 2 p_1 p_2, ~2 p_2^2 } \right )^{-1}
  \left ( \matrix { B_1^{(13)}-B_1^{(23)}+f_1 C_{12}^{(123)} \cr
 -B_1^{(13)}+f_2 C_{12}^{(123)}-2 C_{24}^{(123)}  } \right ) ~~,
\eqa
\bqa
C_{24}^{(124)} &=& \frac{1}{4}+\frac{1}{4} B_0^{(24)}+\frac{m_1^2}{2} C_0^{(124)}
-\frac{f_1}{4} C_{11}^{(124)}-\frac{f_2+f_3}{4} C_{12}^{(124)} ~~, \nonumber \\[.3cm]
\left ( \matrix{ C_{21}^{(124)} \cr C_{23}^{(124)} }\right )
 &=& \left ( \matrix{ 2 p_1^2 , ~~ 2 p_1 (p_2+p_3)  \cr
  2 p_1 (p_2+p_3),~ 2 (p_2+p_3)^2 } \right )^{-1}
 \left ( \matrix { B_1^{(14)}+B_0^{(24)}+f_1 C_{11}^{(124)}-2C_{24}^{(124)} \cr
 B_1^{(12)}-B_1^{(14)}+(f_2+f_3) C_{11}^{(124)}  } \right )
~~,  \nonumber \\[.3cm]
 \left ( \matrix{ C_{23}^{(124)} \cr C_{22}^{(124)} }\right )
 &=& \left ( \matrix{ 2 p_1^2 , ~~ 2 p_1( p_2+p_3)  \cr
 2 p_1 (p_2+p_3),~ 2 (p_2+p_3)^2 } \right )^{-1}
 \left ( \matrix { B_1^{(14)}-B_1^{(24)}+f_1 C_{12}^{(124)} \cr
 -B_1^{(14)}+(f_2+f_3) C_{12}^{(124)}-2 C_{24}^{(124)}  } \right ) ~~,\nonumber \\
\eqa
\bqa
C_{24}^{(134)} &=& \frac{1}{4}+\frac{1}{4} B_0^{(34)}+\frac{m_1^2}{2} C_0^{(134)}
-\frac{(f_1+f_2)}{4} C_{11}^{(134)}-\frac{f_3}{4} C_{12}^{(134)} ~~, \nonumber \\[.3cm]
\left ( \matrix{ C_{21}^{(134)} \cr C_{23}^{(134)} }\right )
 &=& \left ( \matrix{ 2 (p_1+p_2)^2 ,~ 2 (p_1+ p_2)p_3  \cr
 2 (p_1+ p_2)p_3 ,~ ~ 2 p_3^2 } \right )^{-1}
  \left ( \matrix { B_1^{(14)}+B_0^{(34)}+(f_1+f_2) C_{11}^{(134)}-2C_{24}^{(134)} \cr
 B_1^{(13)}-B_1^{(14)}+f_3 C_{11}^{(134)}  } \right )
~~, \nonumber  \\[.3cm]
 \left ( \matrix{ C_{23}^{(134)} \cr C_{22}^{(134)} }\right )
 &=& \left ( \matrix{ 2 (p_1+p_2)^2 ,~ 2 (p_1+ p_2)p_3  \cr
 2 (p_1+ p_2)p_3 , ~~2 p_3^2 } \right )^{-1}
  \left ( \matrix { B_1^{(14)}-B_1^{(34)}+(f_1+f_2) C_{12}^{(134)} \cr
 -B_1^{(14)}+f_3 C_{12}^{(134)}-2 C_{24}^{(134)}  } \right ) ~~,
\eqa
\bqa
C_{24}^{(234)} &=& \frac{1}{4}+\frac{1}{4} B_0^{(34)}+\frac{m_2^2}{2} C_0^{(234)}
-\frac{(f_2+2 p_1p_2)}{4} C_{11}^{(234)}-\frac{(f_3+2 p_1 p_3)}{4} C_{12}^{(234)} ~~, \nonumber \\[.3cm]
\left ( \matrix{ C_{21}^{(234)} \cr C_{23}^{(234)} }\right )
 &=& \left ( \matrix{ 2 p_2^2 ,~ 2  p_2p_3  \cr
 2  p_2 p_3 ,~ ~ 2 p_3^2 } \right )^{-1}
  \left ( \matrix { B_1^{(24)}+B_0^{(34)}+(f_2+2p_1p_2) C_{11}^{(234)}-2C_{24}^{(234)} \cr
 B_1^{(23)}-B_1^{(24)}+(f_3+2p_1p_3) C_{11}^{(234)}  } \right )
~~, \nonumber  \\[.3cm]
 \left ( \matrix{ C_{23}^{(234)} \cr C_{22}^{(234)} }\right )
 &=& \left ( \matrix{ 2 p_2^2 ,~ 2  p_2p_3  \cr
 2  p_2 p_3 , ~~2 p_3^2 } \right )^{-1}
  \left ( \matrix { B_1^{(24)}-B_1^{(34)}+(f_2+2p_1p_2) C_{12}^{(234)} \cr
 -B_1^{(24)}+(f_3+2p_1p_3) C_{12}^{(234)}-2 C_{24}^{(234)}  } \right ) ~~.
\eqa


%\clearpage
%\newpage

\begin{figure}
\centering
\epsfig{file=D0.eps, width=16cm, angle=0}
\vspace{1.5cm}
\caption{Comparison of the exact  and asymptotic results for
$s^2D_0$, as a function of $\sqrt{s}/M$, ~ at ~ $m_i^2 = p_j^2 = M^2$
and $t=-s/2$. Real and Imaginary parts are studied in the
left and right panels respectively. }
\label{fig:d0}
\end{figure}

\end{document}